# FINANCIAL VIABILITY OF SOCIAL ENTERPRISES

**Zoltan Bartha**

**Adam Bereczk**

University of Miskolc, **Hungary**

**ABSTRACT**

Our study presents a model of factors influencing the financial viability of Hungarian social enterprises, and tests the model on a sample of 220 Hungarian firms involved in social entrepreneurship. In the model we suggest that the most important factors for financial viability are the Regulatory environment (the transparency of regulations); the Entrepreneurial attributes of the entrepreneur (business orientation, business skills and experience, business planning tendencies); the Financial support provided by the environment (the ratio of grants, donations and subsidies within the total revenues of the firm); and the Strategy followed by the firms (the presence of such generic strategies as cost leadership or differentiation). We find that only two of the model's four factors are significantly associated with Financial viability: Entrepreneurial attributes and Financial support. The results suggest that the best way of strengthening the viability of social enterprises is through entrepreneurship training (to enhance the business skills and experience of the entrepreneurs, and to propagate business planning), and to provide grants and subsidies to these firms. As no significant association was found between Financial viability and Strategy, we can conclude that the role of market competition is probably relatively week among Hungarian social enterprises.

## INTRODUCTION

The Entrepreneurship Research Group of the University of Hohenheim published its report titled "Most promising topical areas in entrepreneurship research" in 2018, which is based on feedback received from 225 experienced entrepreneurship researchers [1]. Social entrepreneurship is featured on top of the list, in the third place, higher than such trending topics like entrepreneurship psychology or innovation in entrepreneurships. Within the topic of social entrepreneurships, two of the most important research areas are financial viability and financial self-sufficiency. Social enterprises are crucial for long term economic development, but very often they rely on revenues that are outside of the scope of their business activity, which on the other hand threatens their long term financial sustainability. In this paper we develop a model for the financial viability (financial sustainability and stability) of social enterprises, and uncover the factors that have the significant influence on the viability.

## THEORETICAL BACKGROUND

In the area of financial viability and financial self-sufficiency, there have emerged two distinct schools of social entrepreneurship. The difference between the two schools is rooted in the dilemma of social innovation vs. market revenues. Defourny & Nyssen [2], and Dees [3] belong to the social innovation school. They suggest that the social mission is the primary principle in social enterprises. As a result they propagate that enterprises may rely on a wide range of funding sources (from market revenues to donations) in order to achieve social change. The earned income school on the other hand stresses the entrepreneurial nature of social enterprises. They believe that revenues earned in the market is what makes a social enterprise a true entrepreneurship.

The 2012 study by Abu-Saifan titled Social Entrepreneurship: Definitions and Boundaries [4] gives an overview on the definitions created for social entrepreneurs and enterprises, and suggests a "final" definition of the term. Abu-Saifan believes that the essential characteristics of the social enterprises are independency, self-sufficiency, and sustainability: "the social entrepreneur is a mission-driven individual who uses a set of entrepreneurial behaviours to deliver a social value to the less privileged, all through an entrepreneurially oriented entity that is financially independent, self-sufficient, or sustainable" [4, pp 25].

Madill [5] also agrees that the earned income and the financial self-sufficiency and sustainability are the essential aspects of social enterprises. She also remarks that some authors draw a division line between sustainability and self-sufficiency: "sustainability can be achieved through philanthropy, donations, grants, government subsidy and earned income, but self sufficiency can only be achieved through reliance on earned income" (p. 120).

Based on the above we apply the following approach in our study:

- accepting the primary nature of the social mission (highlighted by the social innovation school), we use the term financial viability to describe the financial position and structure of social enterprises (as opposed to measuring their financial performance as the mere difference between their revenues and expenditures);
- social enterprises can have different origins and forms (from subsidiaries of non-profit organisations to social purpose business ventures [6]), so we regard independency as an issue not related to financial factors;
- based on the earned income approach we regard it as a higher level of financial viability if the financial sustainability is achieved at a higher self-sufficiency level.

There are large regional differences in these social enterprise criteria. According to Madill et al. [7], around half of the Canadian social enterprises has achieved a high level of financial self-sufficiency. In Hungary on the other hand typically the earned income is way below the expenditures and so social enterprises have to rely on other forms of revenues (such as grants, donations and government subsidy) [8]. In the next section we present the model developed to measure the self-sufficiency of social enterprises, and test in on a Hungarian sample.

**CONCEPTUAL FRAMEWORK**

Our model is based on the one presented by Saebi et al. in 2018 [9] in the "A Framework for Identifying and Organising Research Opportunities in SE" section of their paper. They created their model based on a structure review of the social enterprises literature, with the purpose to point out the essential research areas of the field. Saebi et al. list nine crucial areas, show the possible relationships among them on three possible levels (macro, meso and micro), and two different stages (pre-formation and post-formation stage). We did not incorporate all the aspects of the model due to the lack of proper measurement methods. Our REAFS model of financial viability includes at least an aspect from each of Saebi et al.'s levels (see Figure 1):

- (R) Regulatory environment. A macro-level component that reflects the institutional context. Based on Gwendolyn & Lauritzen [10] we measure R as the perceived transparency of regulations.

- (F) Financial support & (S) Strategy. One of the important meso-level factors in Saebi et al.'s model is the social enterprise (as an organisation). We concentrate on the resources (resource-based view), and the governance and management factors. As Hungarian social enterprises are notoriously underfinanced, from the former group (resources) we focus on financial resources – (S) Financial support. Based on Porter [11] the governance and management factor is measured with the presence of such generic strategies as cost leadership or differentiation.

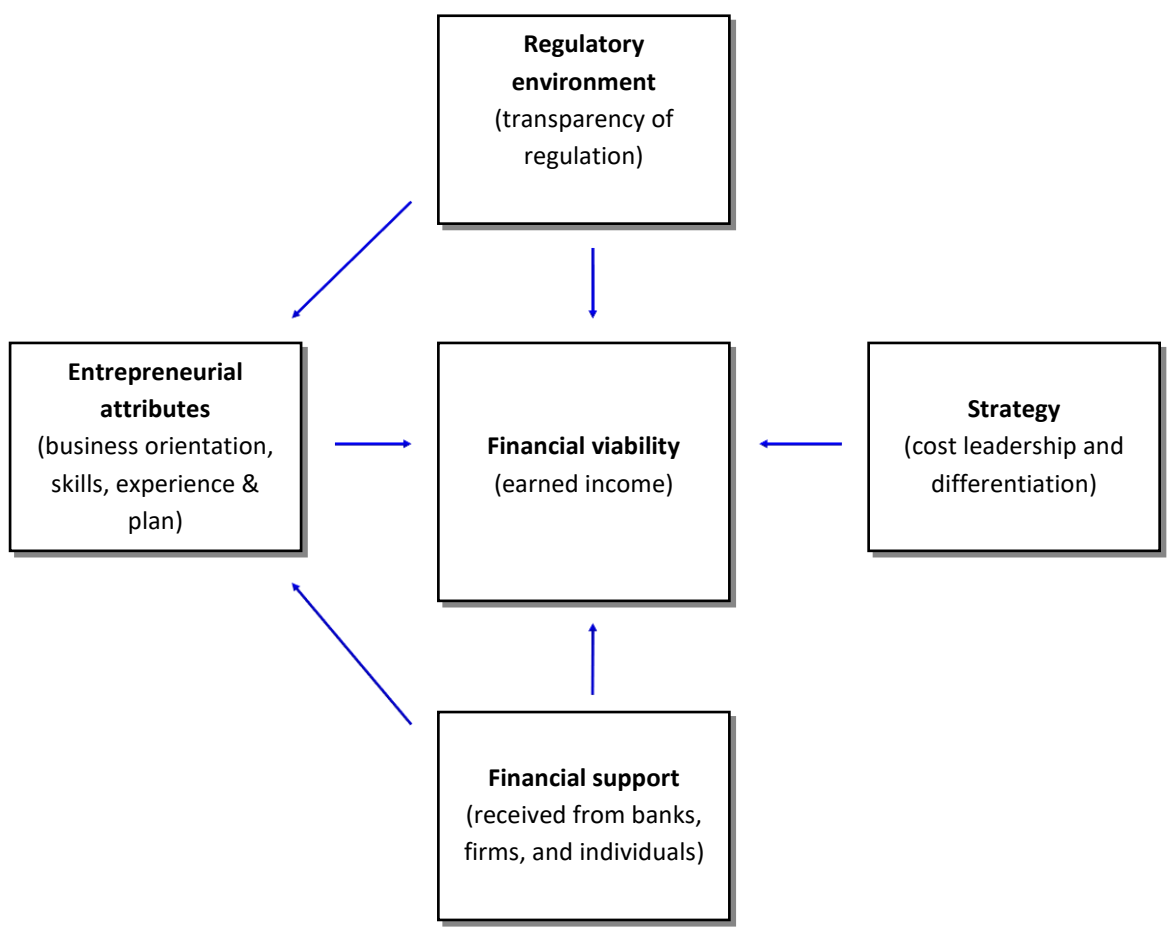

**Figure 1. The REAFS model of social enterprises' financial viability**

- (EA) Entrepreneurial attributes. From the micro-level components we concentrate on Saebi et al.'s social entrepreneur factor. It has been shown that the skills and attributes of the entrepreneur are essential in the success of social entrepreneurships. Urbano [12] reviews a large chunk of the social entrepreneurship literature and concludes that articles published in top journals (listed in the Web of Science) tended to focus on describing the experiences of the most popular social entrepreneurs. We measure the Entrepreneurial attributes of the social entrepreneur with the intensity of business orientation, the level of business skills, entrepreneurial experience, and experience in business planning. Dees et al. [13] also stress the importance of the latter factor when they note that the opposition of social entrepreneurs to business planning is probably due to the fact that they only see it as a bureaucratic function. A change in this attitude is highly desirable, according to Dees et al.

**ANALYSIS**

The data for this study come from a survey that was conducted by the Faculty of Economics, University of Miskolc in 2017 by a team lead by Éva G. Fekete and Ágnes Horváth Kádárné [8]. The survey was sponsored by the Hungarian Employment Agency OFA, within the project called PiacTárs – Priority project for the support of social enterprises, and for the creation of a sustainable and competitive social economy. A total 220 social enterprises were surveyed.

The factors of the REAFS model were measured the following way.

- (R) Regulatory environment: variable with three values – +1 if the transparency of the regulations is viewed as an opportunity for the firm by the entrepreneur; -1 if the lack of transparency is viewed as a threat to the firm by the entrepreneur; and 0, if it not regarded neither as an opportunity nor as a threat.
- (EA) Entrepreneurial attributes is calculated as a sum of four variables measuring the perceived strength and weaknesses of social enterprises, namely:
  - Business orientation, with three possible values: +1 if the business orientation of the firm is viewed as a major strength by the entrepreneur; -1 if the lack of business orientation is viewed as a major weakness; and 0, if it was not mentioned neither as a strength, nor as a weakness.
  - Business skills with three possible values: +1 if the business skills of the entrepreneur is viewed as a major strength by the entrepreneur; -1 if the lack of business skills is viewed as a major weakness; and 0, if it was not mentioned neither as a strength, nor as a weakness.
  - Business experience with three possible values: +1 if the business experience of the entrepreneur is viewed as a major strength by the entrepreneur; -1 if the lack of business experience is viewed as a major weakness; and 0, if it was not mentioned neither as a strength, nor as a weakness.
  - Business planning with three possible value: +1 if the business planning activity of the firm is viewed as a major strength by the entrepreneur; -1 if the lack of business planning activity of the firm is viewed as a major weakness; and 0 if it was not mentioned neither as a strength, nor as a weakness.
- (F) Financial support is measured as a ratio of total support (grants, donations, subsidies) received as a share of total revenues.
- (S) Strategy is calculated as a sum of two variables measuring the perceived strength and weaknesses of social enterprises, namely:

- o Cost leadership with three possible values: +1 if the product/service price is viewed as a major strength by the entrepreneur; -1 if the product/service price is viewed as a major weakness; and 0, if it was not mentioned neither as a strength, nor as a weakness.
  - o Differentiation with three possible values: +1 if the uniqueness of the product/service is viewed as a major strength by the entrepreneur; -1 if the uniqueness of the product/service is viewed as a major weakness; and 0, if it was not mentioned neither as a strength, nor as a weakness.
- (FV) Financial viability is measured as a ratio of earned income (market revues) as a share of total revenues.

Basic association relationships (Eta squared for testing the relationship between a variable measured on ordinal level, and one, on ratio level; Cramer's V for two variables measured on ordinal level; and Pearson's R for two variables measured on ratio level) were calculated using SPSS. The results are shown in Figure 2. The black arrows represent significant associations/correlations, while the intermittent ones show relationships that should exist, but we found not significant association in our test.

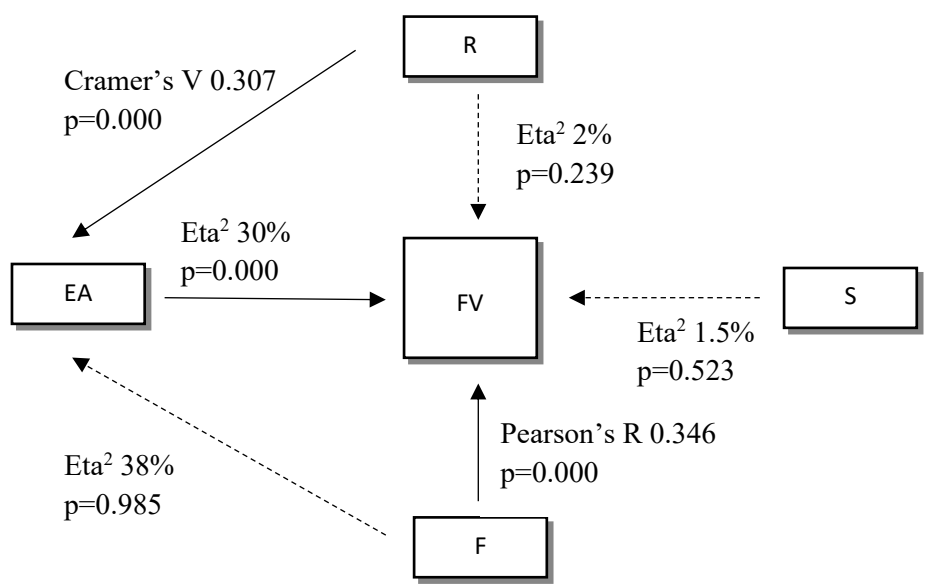

**Figure 2. Test results for the REAFS model.**

Financial viability is significantly associated with the Entrepreneurial attitudes of tested social entrepreneurs (the business orientation, skills and experience of the entrepreneur, as well as the attitude toward business planning). 30% of the variety in the ratio of earned income is explained by this variable. Although we found no direct relationship between the Regulatory environment and the Financial viability, there is a significant association between the Regulatory environment and the Entrepreneurial attributes (Cramer's V = 0.307). Those who see the lack of regulatory transparency as a threat, tend to be more entrepreneurial oriented, skilled and experienced.

The other factor being significantly associated with Financial viability, is Financial support (Pearson's R = 0.346). The results suggests that firm that receive a larger ratio of their revenues as grant, donations and subsidies, are tend to be more financially viable as well.

The Strategy is not significantly associated with the Financial viability of the firm, and the Financial support is not significantly associated to the Entrepreneurial attributes either.

**CONCLUSION**

Social enterprises play a crucial role in long term economic development, but their financial viability, especially in Hungary and in the Central-Eastern European region is relatively weak. We developed the REAFS model to determine the most important influencing factors of financial viability. Based on the literature review we assumed that the transparent regulatory environment (R – Regulatory environment); better business orientation, business skills and experience, and thorough business planning (a four together formed the EA – Entrepreneurial attributes variable); more financial support (S – Financial support; and the presence of a generic strategy (either cost leadership or differentiation; S – Strategy) positively affect the viability of social enterprises.

Out of the four possible factors, two could have a significant impact on the financial viability: Entrepreneurial attributes and Financial support. A more skilled and experienced entrepreneur, in an environment that can provide significant donations and subsidies can enhance the financial viability of the social enterprise. Transparent regulations are only significantly associated with the Entrepreneurial attributes, and so do not seem to directly affect the viability of the enterprises. There is no significant relationship between the presence of a generic strategy, and the financial viability either. This may suggest that the level of market competition is fairly low among Hungarian social enterprises.

Our results suggest that policy should focus on enhancing the entrepreneurial skills and experience of social entrepreneurs, point to the importance of business planning, and also highlight that financial support (in form of grants, donations and subsidies) seem to be crucial in terms of financial viability in Hungary. One major limitation of our test is the low (nominal and ordinal) level of measurement in case of most of our factors, as well as the low level of reporting of revenues and different revenue categories, which lead to missing values in case of many of the 220 surveyed firms. A more reliable test could be conducted if the measurement level of our variables was further developed, and if more respondents reported certain ratios about their revenues.

**ACKNOWLEDGEMENTS**

This research was supported by the project nr. EFOP-3.6.2-16-2017-00007, titled Aspects on the development of intelligent, sustainable and inclusive society: social, technological, innovation networks in employment and digital economy. The project has been supported by the European Union, co-financed by the European Social Fund and the budget of Hungary.